\documentclass[aps,pra,twocolumn,showpacs,floatfix,superscriptaddress]{revtex4}
\usepackage{amsmath,amssymb,amsfonts,bbm,graphicx,hyperref,times}
\def\PDD{P_{\rm K}}
\def\PHD{P_{\rm H}}
\def\PQQ{P_{\rm Q}}
\def\hvarrho{\hat{\varrho}}
\def\ha{\hat{a}}
\def\hU{\hat{U}}

\def\hPi{\hat{\Pi}}
\def\hx{\hat{x}}
\begin{document}
%%%%%%%%%%%%%%%%%%%%%%%%%%%%%%%%%
%\title{Discrimination of phase-shifted coherent signals in the
%  presence of phase diffusion}
\title{Homodyne detection as a near-optimum receiver for phase-shift 
keyed binary communication in the presence of phase diffusion}
\author{Stefano Olivares}
\affiliation{Dipartimento di Fisica, Universit\`a degli Studi di
  Milano, I-20133 Milano, Italy} \affiliation{CNISM, UdR Milano
  Statale, I-20133 Milano, Italy.}
\author{Simone Cialdi}
\affiliation{Dipartimento di Fisica, Universit\`a degli Studi di
  Milano, I-20133 Milano, Italy} \affiliation{INFN, Sezione di Milano,
  I-20133 Milano, Italy}
\author{Fabrizio Castelli}
\affiliation{Dipartimento di Fisica, Universit\`a degli Studi di
  Milano, I-20133 Milano, Italy} \affiliation{INFN, Sezione di Milano,
  I-20133 Milano, Italy}
\author{Matteo G. A. Paris}
\affiliation{Dipartimento di Fisica, Universit\`a degli Studi di
  Milano, I-20133 Milano, Italy} \affiliation{CNISM, UdR Milano
  Statale, I-20133 Milano, Italy.}
\date{\today}
%%%%%%%%%%%%%%%%%%%%%%%%%%%%%%%%%
\begin{abstract}
  We address binary optical communication channels based on
  phase-shift keyed coherent signals in the presence of phase
  diffusion. We prove theoretically and demonstrate experimentally
  that a discrimination strategy based on homodyne detection is robust
  against this kind of noise {for any value of the channel
    energy}. Moreover, we find that homodyne receiver beats the
  performance of Kennedy receiver as the signal energy increases, and
  achieves the Helstrom bound in the limit of large noise.
\end{abstract}
\maketitle
%%%%%%%%%%%%%%%%%%%%%%%%%%%%%%%%%
{\em Introduction} --- In a binary quantum communication channel the
sender encodes the logical symbols ``1'' and ``0'' on two states of a
physical system described by the density operators $\hvarrho_1$ and
$\hvarrho_0$, respectively. In order to retrieve the logical
information, the receiver should discriminate between the two signals
which, in general, may be not orthogonal, due to the encoding process
itself ($\hvarrho_1$ and $\hvarrho_0$ may refer to non-orthogonal
states) or because of the noise during the propagation stage
\cite{hel:76}. In this case, an unavoidable probability of error
$P_e=\frac12[P(1|0)+P(0|1)]$ appears, where $P(j|k)$ is the
probability of inferring the symbol $j$ when the signal sent over the
channel was meant for $k$, and where we assumed that the two states
are sent with equal probability.  The minimum error probability
allowed by quantum mechanics is given by the Helstrom bound
\cite{hel:76} $P_Q = \frac12(1-\frac12\, {\rm Tr} | \hvarrho_0 -
\hvarrho_1|)$,
where $|A|=\sqrt{A^\dag A}$. Optimal discrimination of non orthogonal
states is a crucial topic for the effective implementation of quantum
communication channels and, as a consequence,  different strategies has been employed to
attack the problem in different situations \cite{r1,r2,r3}.
\par
In the following, we address binary communication channels with
phase-shift keyed (PSK) signals \cite{Hir96,Hir99}, i.e., channels
where the information is encoded on two coherent states $|\psi_1
\rangle=|\alpha\rangle$ and $|\psi_0 \rangle=|-\alpha\rangle$ (without
lack of generality we can assume $\alpha \in {\mathbbm R}$, $\alpha
>0$). In this case, the Helstrom bound rewrites
as: $P_Q = \frac12\left[1-\sqrt{1-|\langle \psi_0|
\psi_1\rangle|^2}\right]$, i.e.
\begin{align}\label{pe:pure}
P_Q = \frac12\left[1-\sqrt{1-\exp{(-4N)}}\right],
\end{align}
where $N=|\alpha|^2$ is the average number of photons
contained in each signal, and it will be referred to as the {\em signal
energy} throughout the paper.
A detection strategy  achieving this level of error probability is said to be an 
{\em optimum receiver}. 
\par
Quantum state discrimination strategies for PSK signals is an active
field of research since it encompasses the read-out problem, and it is
an important tool to assess the performances of quantum limited
measurements in optical communication, e.g., for deep-space missions
\cite{lau:06}.  On the one hand, {in the high energy regime,} a
near-optimum receiver based on photodetection, the so-called Kennedy
receiver, has been proposed long ago for coherent signals propagating
in an ideal channel \cite{ken:73,tak03}.  Kennedy receiver has been
also extended \cite{dol:73,ass11,vil12}, and, in particular, an
optimum receiver based on an adaptive scheme has been recently
experimentally realized \cite{coo:07}.  On the other hand, strategies
based on homodyne detection have great practical advantages
\cite{gre12}, and for this reason their performances have been
analyzed in realistic situations.  In particular, receivers based on
homodyne detection has been theoretically \cite{ban97,oli:04,car11}
and experimentally investigated \cite{witt:10} in the presence of
losses and phase insensitive thermal noise. More generally, these
receivers have been proved to represent the optimum Gaussian strategy
for the discrimination of the PSK coherent signals \cite{tak08}, and
it has been also demonstrated that it possible to emulate adaptive
processes by means of postprocessing and Bayesian analyses
\cite{Mig11}.  Hybrid \cite{Mul12} and displacement-based \cite{Isu12}
detectors have been also implemented to decrease the error
probability, with application to $M$-ary communication channels
\cite{Nai12}.
\par
{Communication schemes based on coherent signals may be useful in
  scenarios in which quantum resources (as single photons and
  entanglement) cannot be fully exploited, as in free-space
  communication. In this paper we analyze a source of noise, namely,
  phase diffusion, which is detrimental for coherent-signal based
  channels. At the same time, this provides a relevant example of
  non-Gaussian noise. } We demonstrate experimentally the robustness
of the homodyne receiver against phase noise.  Moreover, we find that
in the presence of phase noise, homodyne receiver beats the
performances of Kennedy receiver as the signal energy increases and/or
the noise is larger than an intensity dependent threshold
value. Finally, we show that homodyne detection achieves the Helstrom
bound in the limit of large noise.
\par
\begin{figure}[h!]
\includegraphics[width=0.49\columnwidth]{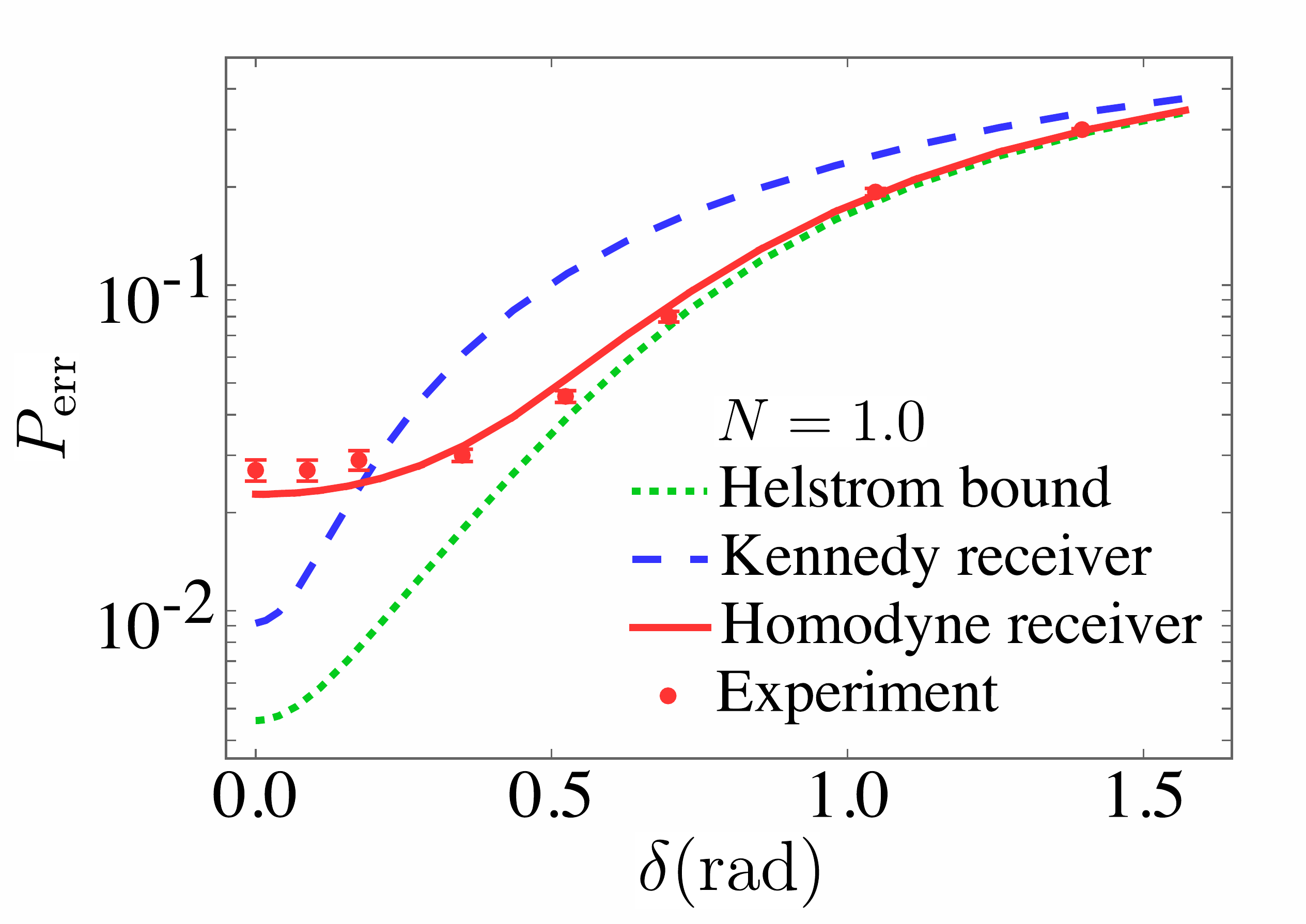}
\includegraphics[width=0.49\columnwidth]{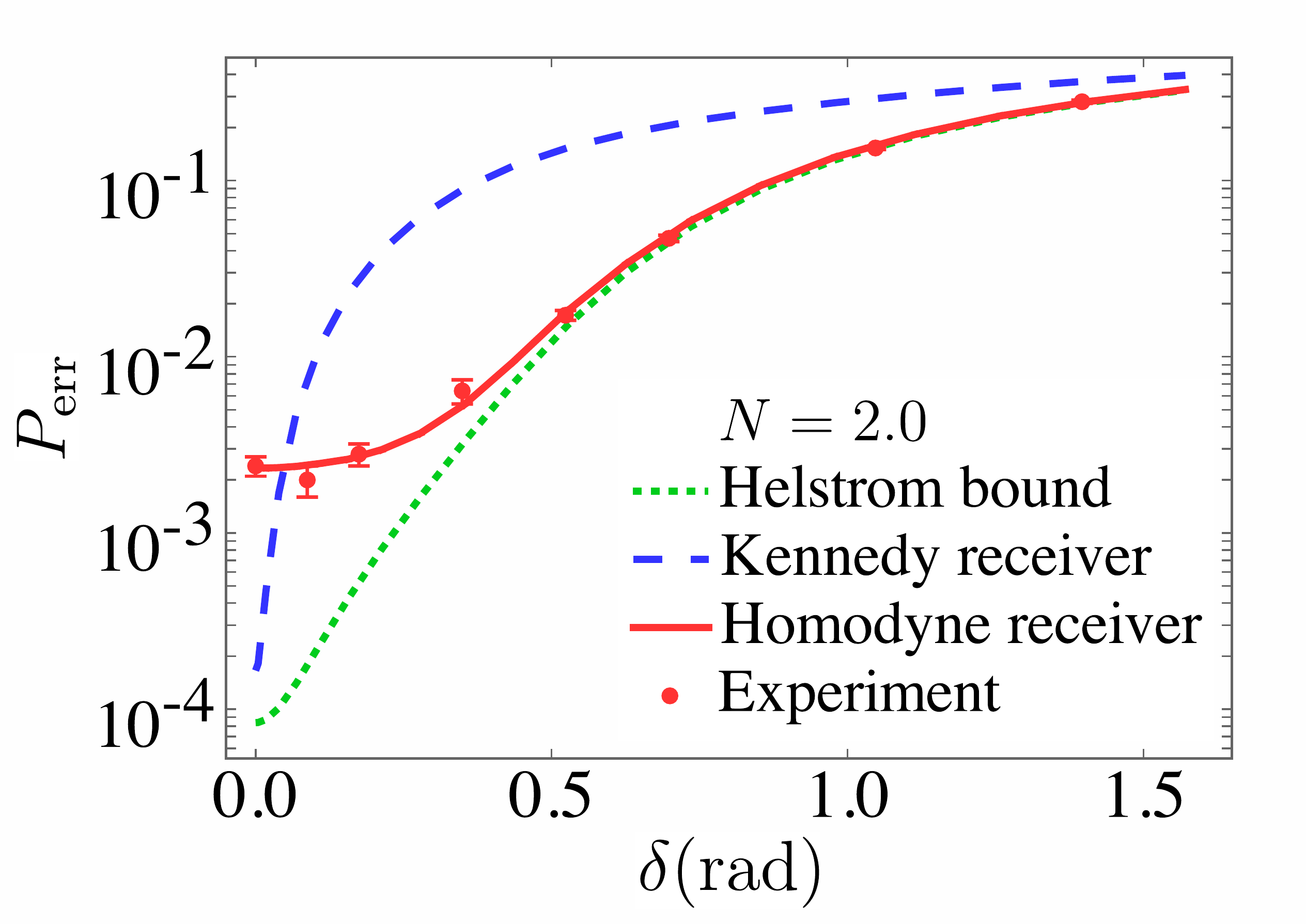}
\includegraphics[width=0.49\columnwidth]{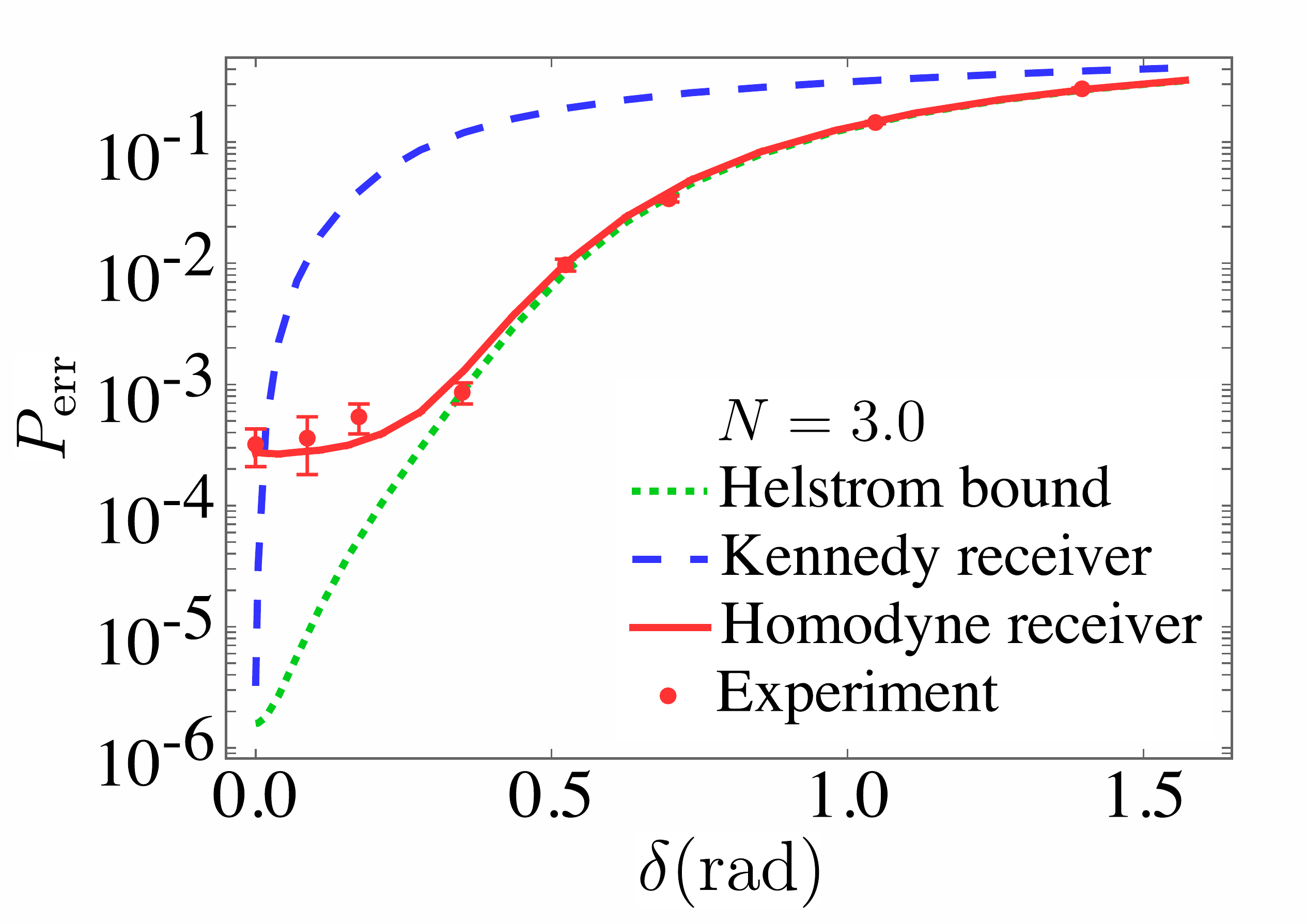}
\includegraphics[width=0.49\columnwidth]{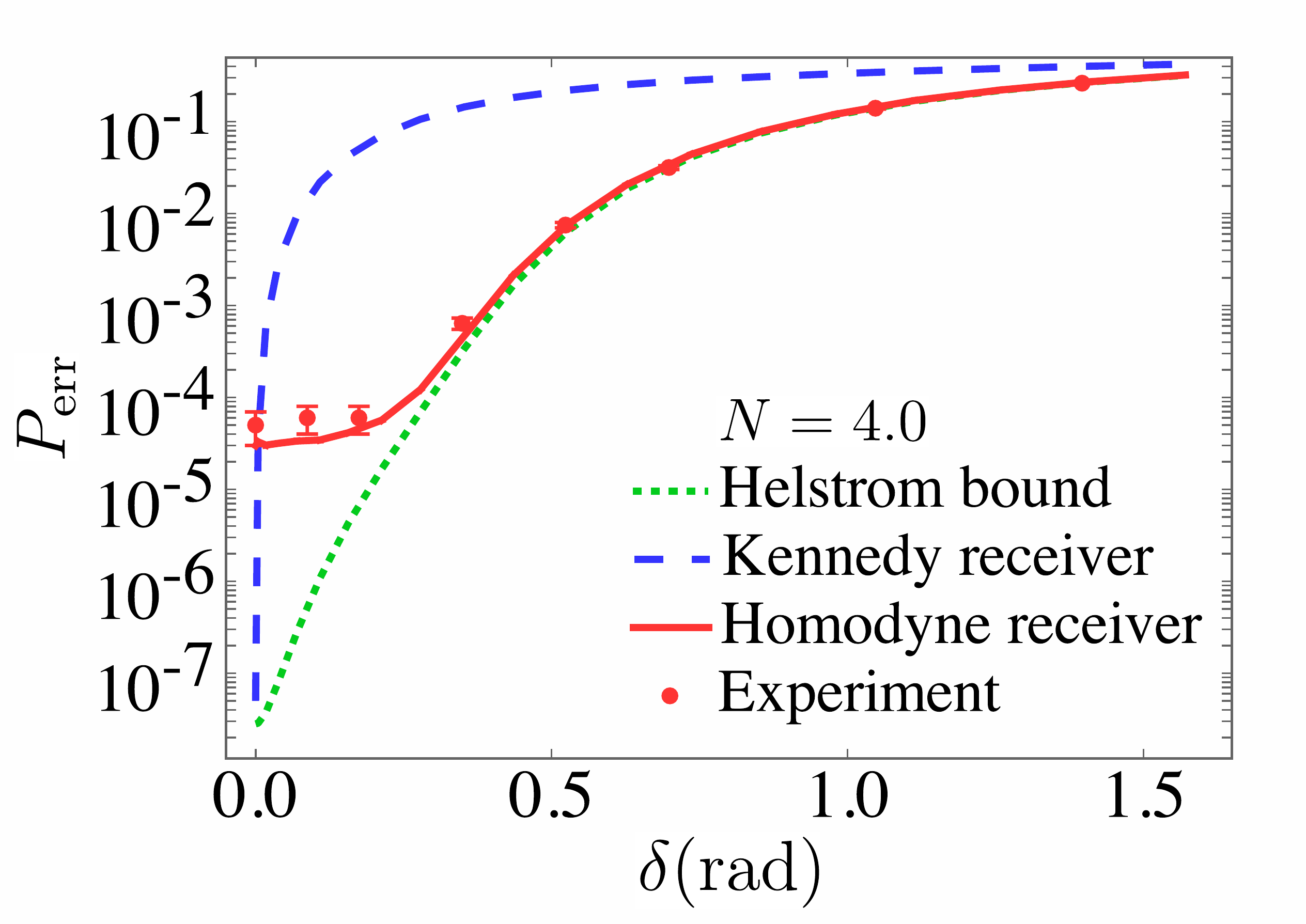}
\caption{(Color online) Log-linear plot of the homodyne probability of 
error (red circles) as a function of the diffusion parameter $\delta$ 
for different values of the signal energy $N$.  Each point corresponds
to the average of ten acquisitions of $5\times 10^3$ values 
with a phase diffusion frequency of 1~kHz. The errors bars corresponds to the
standard deviation of the mean.  The theoretical predictions 
$P_H(\delta)$ (solid red) is reported for comparison. We also show the error 
probability $P_K(\delta)$ of a Kennedy receiver (dashed blue) in the same 
experimental 
conditions and the Helstrom bound $P_Q(\delta)$ (dotted green).
\label{f:exp}}
\end{figure}

{\em Phase diffusion} --- {Any phase diffusion process coming from a
non-dissipative interaction with a bosonic environment may be
described by a suitable phase-diffusion Master equation \cite{GardQN}.} The
overall effect of phase diffusion on a coherent state
$\hvarrho_{\alpha} =| \alpha \rangle \langle \alpha |$ is described by
the map \cite{geno:11}:
\begin{align}
\hvarrho_{\alpha} \to \hvarrho'_{\alpha} = 
\int_{\mathbbm R} d\phi\, g(\phi,\delta)\,
| \alpha\, e^{i\phi} \rangle \langle \alpha\, e^{i\phi} |,
\end{align}
where $ g(\phi,\delta)$ is a normal distribution of the variable
$\phi$ with zero mean and standard deviation $\delta$.
At the end of a channel affected by phase
diffusion the receiver is faced with the problem of discriminating
between $\hvarrho'_\alpha$ and $\hvarrho'_{-\alpha}$. 
In order to calculate the corresponding
Helstrom bound, one has to diagonalize the traceless operator
$\Lambda=\hvarrho'_{\alpha} - \hvarrho'_{-\alpha}$ whose expansion in the
photon-number basis reads:
\begin{align}
\Lambda =\sum_{n,m=0}^{\infty}
\frac{\alpha^{n+m}}{\sqrt{n! \, m!}}\,
e^{-\alpha^2-\frac{(n-m)^2}{2}\delta^2}
[1 - (-1)^{n-m}]\,
| m \rangle \langle n |\nonumber \,.
\end{align}
This can be done numerically after a suitable truncation of the 
Hilbert space, which is determined by the average number of 
photons of the two signals. The Helstrom bound for different 
values of the signal
energy $N$  and the diffusion coefficient $\delta$ is reported
in Fig. \ref{f:exp}, together with the experimental data for the
homodyne receiver. 
For small values of the coherent amplitude we may truncate $\Lambda$ at
low dimension achieving the analytic expression
\begin{align}
\PQQ (\delta) \simeq \frac12 \left(1- \alpha e^{-\frac12
\delta^2}\right)\,, \label{asyQA}
\end{align}
whereas, in the limit $\delta\gg 1$ of large noise we have
\begin{align}
\PQQ (\delta) \simeq \frac12 \left(1-g_{\rm Q}(\alpha)\, e^{-\frac12
\delta^2}\right)\,, \label{asyQD} 
\end{align}
where $g_{\rm Q}(\alpha)$ is a decreasing function of the amplitude.
\par
{\em Kennedy receiver in the presence of phase noise} --- In the ideal
case, a near-optimum strategy is provided by the Kennedy receiver
\cite{ken:73} based on photodetection: the signal mode $\ha$ excited
either in $|\alpha\rangle$ or $|-\alpha\rangle$, interferes with a
reference mode $\hat{b}$ excited in the coherent state $|\beta\rangle$
at a beam splitter (BS) with transmissivity $\tau=\cos^2\varphi$ and a
photodetector detects the light at one output. The overall evolved
state reads $U_\tau |\pm \alpha \rangle \otimes | \beta \rangle = |\pm
\sqrt{\tau}\alpha + \sqrt{1-\tau}\beta \rangle \otimes
|\sqrt{\tau}\beta \mp \sqrt{1-\tau}\alpha \rangle$, where $\hU_{\tau}$
is the evolution operator associated with the BS.  If we choose
$\beta=\alpha\sqrt{\tau}/\sqrt{1-\tau}$, take $\tau\to 1$ and
consider only the first output port, then we have the following input-output
relation $|\pm\alpha\rangle \to |\pm \alpha+\alpha\rangle$. Therefore, a
natural discrimination strategy is to associate the absence of light
(the vacuum) with the symbol ``0'', i.e., $|-\alpha\rangle$, and the
detection of any number of photons with the symbol is ``1'', i.e.,
$|\alpha\rangle$. This strategy also implies a nonzero probability of
inferring the wrong symbol, which is determined by the conditional
probabilities \cite{oli:04}:
\begin{subequations}\label{P:cond:DD}
\begin{align}
P(0|1) &= \lim_{\tau\to 1}{\rm Tr}[\hU_{\tau}\hvarrho_{\alpha}\otimes \hvarrho_{\beta}
\hU_{\tau}^{\dag}\, \hPi_0\otimes\hat{\mathbbm I}] = e^{-4N}, \\
P(1|0) &= \lim_{\tau\to 1} 
{\rm Tr}[\hU_{\tau}\hvarrho_{-\alpha}\otimes \hvarrho_{\beta}\hU_{\tau}^{\dag} \, 
\hPi_1\otimes\hat{\mathbbm I}]
=0,
\end{align}
\end{subequations}
where we introduced the positive-operator valued measure $\hPi_0 = | 0
\rangle\langle 0 |$ and $\hPi_1 = \hat{\mathbbm I} - \hPi_0$,
describing an on-off detector detecting the absence or the presence of
photons, respectively.  The overall probability of error (still
assuming the two signals sent with the same probability) is:
\begin{equation}
\PDD = \frac{P(0|1) + P(1|0) }{2} = \frac{\exp(-4N)}{2}.
\end{equation}
It is worth noting that, in the limit $N\gg 1$,
 $\PDD \approx 2 P_Q$, i.e., the Kennedy receiver based on 
 photodetection is nearly optimal. In the
presence of phase diffusion, Eqs.~(\ref{P:cond:DD}) become:
\begin{subequations}\label{P:cond:DD:phn}
\begin{align}
P_\delta (0|1) &= \int_{\mathbbm R} d\phi\, g(\phi,\delta)\,
\exp\left[ -4\alpha^2\cos^2(\phi/2) \right],\\
P_\delta (1|0) &= 1- \int_{\mathbbm R} d\phi\, g(\phi,\delta)\,
\exp\left[ -4\alpha^2\sin^2(\phi/2) \right],
\end{align}
\end{subequations}
and 
$$\PDD(\delta) = \frac12 \left[ P_\delta(0|1) + P_\delta(1|0)
\right]\,,$$
which can be easily evaluated numerically.
For small values of the coherent amplitude 
we have the analytic expression
\begin{align}
\PDD (\delta) \simeq \frac12 \left(1- 4 \alpha^2 e^{-2
\delta^2}\right)\,, \label{asyKA}
\end{align}
whereas, in the limit $\delta\gg 1$ of large noise we have
\begin{align}
\PDD (\delta) \simeq \frac12 \left(1-g_{\rm K}(\alpha)\, e^{-2
\delta^2}\right)\,, \label{asyKD} 
\end{align}
where $g_{\rm K}(\alpha)$ is a decreasing function of the amplitude.
\par
{\em Homodyne receiver in the presence of phase noise} --- 
Let us now focus on a different strategy based on homodyne
detection: the receiver measures the quadrature $\hx_\psi =
2^{-1/2}(\ha^{\dag}e^{i\psi}+\ha\, e^{-i\psi})$, where $\psi =
\arg[\alpha]$ (in our case $\psi = 0$ or $\psi=\pi$) and associates 
the symbol ``1'' (``0'') to a positive
(``negative'') outcome. In the presence of losses and thermal noise
but without phase noise, this strategy has been proven to beat the
photodetection strategy either in the low or in the high energy 
regime \cite{oli:04}. In the presence of phase diffusion we have 
the following conditional probabilities (the noiseless case is
recovered in the limit $\delta\to 0$):
%\begin{subequations}
\begin{align}
\label{P:cond:HD:phn}
Q_\delta (0|1) = \int_{-\infty}^{0}\!\!\!\!\! dx\,
p_\delta(x;\alpha)\,, \;
Q_\delta (1|0) = \int_{0}^{+\infty}\!\!\!\!\!\!\!\! dx\, p_\delta(x;-\alpha),
\end{align}
%\end{subequations}
where:
\begin{equation}
p_\delta(x;\pm\alpha)=\int_{\mathbbm R} \frac{d\phi}{\sqrt{\pi}} \, g(\phi,\delta)\,
e^{-\left( x \mp \sqrt{2} \alpha \cos\phi \right)^2}
\end{equation}
is the homodyne probability, namely, the probability of obtaining as
outcome $x$ addressing the quadrature $\hx_0$ given the input $| \pm
\alpha\rangle$. The overall probability of error thus writes:
\begin{equation}
\PHD(\delta) = \frac12 \left[Q_\delta(0|1) + Q_\delta (1|0)\right]\,,
\end{equation}
and can be easily evaluated numerically.
For small values of the coherent amplitude 
we have the analytic expression
\begin{align}
\PHD (\delta) \simeq \frac12 \left(1- \alpha \sqrt{\frac2\pi}e^{-\frac12
\delta^2}\right)
\label{asyHA}\,, 
\end{align}
whereas, in the limit $\delta\gg 1$ of large noise we have
\begin{align}
\PHD (\delta) \simeq \frac12 \left(1-g_{\rm H}(\alpha)\, e^{-\frac12
\delta^2}\right) 
\label{asyHD}\,, 
\end{align}
where $g_{\rm H}(\alpha)$ is a decreasing function of the amplitude
with $g_{\rm H}(\alpha) <g_{\rm Q}(\alpha)$ and with $g_{\rm
H}(\alpha)$ approaching $g_{\rm Q}(\alpha)$ for increasing amplitude.
\begin{figure}[h!]
\begin{center}
\includegraphics[width=0.9\columnwidth]{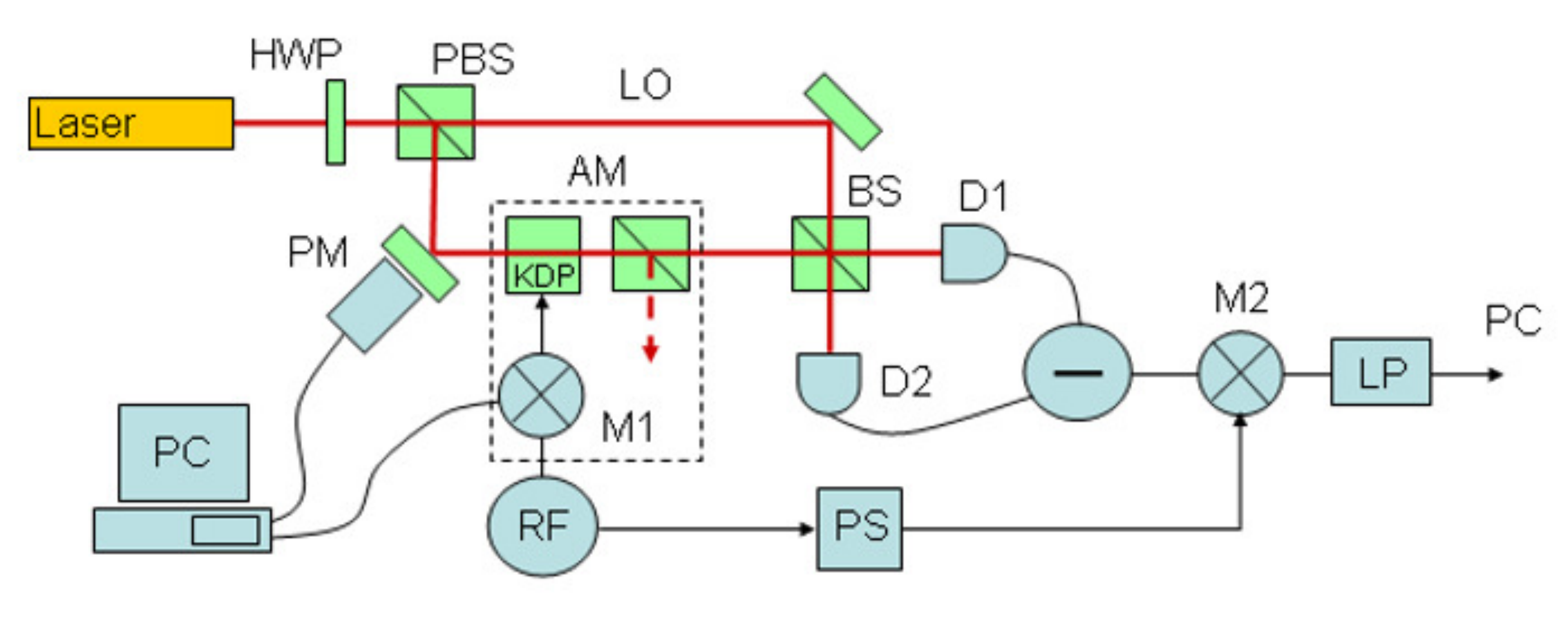}
\end{center}
\caption{(Color online) Schematic diagram of the experimental apparatus.  
The primary source is He:Ne laser, then split into two paths, the local
oscillator (LO) and the signal, by means of a half-wave plate (HWP) 
and a polarizing beam splitter (PBS). The phase modulation (PM) is 
obtained by a piezo connected to a computer (PC). The amplitude modulation 
(AM) of the coherent signals is controlled by a KDP and a PBS. The homodyne 
receiver is composed by a balanced beam splitter (BS) and a balanced amplifier 
detector based on a silicon photodiodes (D1 and D2). M1 is a mixer used to 
set the amplitude modulation by a computer, whereas M2 is a mixer used 
to demodulate the signal @4MHz.
\label{f:scheme}}
\end{figure}
\par
{\em Experimentals} --- 
The experimental apparatus we used to investigate the performance of
the homodyne receiver in the presence of phase diffusion 
is sketched in Fig.~\ref{f:scheme}.
The output beam of a He:Ne laser is split into two paths, the local
oscillator (LO in Fig.~\ref{f:scheme}) and the signal, by means of a
suitable combination of a half wave plate (HWP) and a polarizing beam
splitter (PBS). The phase modulation (PM) is obtained by a piezo
connected to the computer (PC), which sends a voltage signal with a
frequency of 1~kHz corresponding to the phase diffusion. The amplitude
modulation (AM) of the coherent signals is controlled by a suitable
combination of a KDP crystal and a PBS \cite{bri:12}.
The actual homodyne receiver is composed by a 50:50 beam splitter (BS)
and a balanced amplifier detector based on a silicon photodiodes (D1
and D2). In order to experimentally determine the probability of
error, for a fixed value of the signal energy $N = \alpha^2$, we
performed 10 runs with $5\times 10^3$ shots each, randomly chosen
between $| \alpha \rangle$ and $| -\alpha \rangle$.
%%%%%%%%%%%%
\begin{figure}[h!]
\begin{center}
\includegraphics[width=0.9\columnwidth]{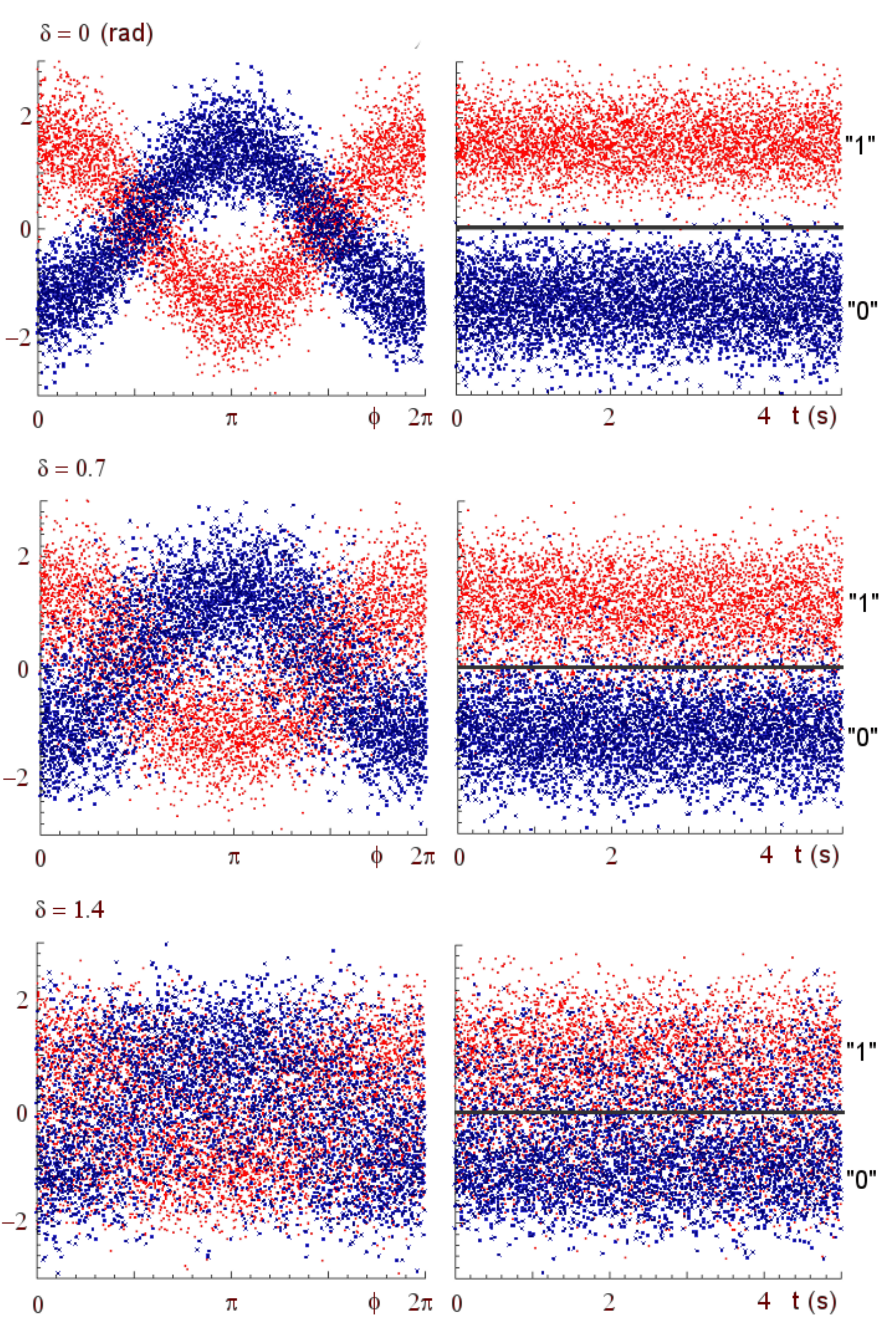}
\end{center}
\caption{(Color online) Left panels: homodyne traces for the quadrature $\hx_{\phi}$,
  of the coherent signals $| \alpha \rangle$ (red) and $| -\alpha
  \rangle$ (blue) for increasing noise (from top to bottom, $\delta =
  0, 0.7, 1.4$~rad) in the case of $N=\alpha^2 = 1.0$. Right panels:
  homodyne traces for the quadrature $\hx_0$ versus the detection
  time. The horizontal line refers to the threshold for the
  discrimination strategy: if the dots fall above (below) that line,
  the signal is chosen to be ``1'' (``0''). \label{f:traces}}
\end{figure}
\par
{\em Data analysis and discussion} --- In Fig.~\ref{f:traces} we
report the data of our experiments for different values of the noise
parameter. In the left panels we show the homodyne traces for the
quadrature $\hx_{\phi}$, of the coherent signals $| \alpha \rangle$
(red) and $| -\alpha \rangle$ (blue) as a function of $\phi$ and for
increasing level of noise. 
In the right panels we report homodyne data for the optimal
quadrature $\hx_0$ versus the detection time. The horizontal line
refers to the threshold for the discrimination strategy: if the dots
(i.e., the homodyne outcomes) fall above (below) that line, the signal is
inferred to be ``1'' (``0'').
\par
In Fig.~\ref{f:exp} we report the behavior of the probabilities of
error $P_Q(\delta)$, $\PDD(\delta)$ and $\PHD(\delta)$ as functions of
the phase diffusion coefficient $\delta$ and for different values of
the input energy $N$. As one may expect, the presence of the phase
noise dramatically affects the performance of the Kennedy receiver: it
is worth noting that as the input energy increases, a small amount of
phase noise in enough to increase of orders of magnitude the
probability of error $\PDD$. On the other hand, the strategy based on
the homodyne detection is quite robust with respect to the phase
noise. As it is apparent from the plots, for $\delta \lesssim 0.2$ the
value of $\PHD$ is almost constant. In addition, $\PHD$ approaches the
quantum mechanical limit given by the Helstrom bound $P_Q$ as $\delta$
increases, as anticipated by the asymptotic expansions in Eqs.
(\ref{asyQD}) and (\ref{asyHD}). In general, as the signal energy and the 
noise increase, the homodyne receiver becomes more effective than 
Kennedy receiver.
\par
\begin{figure}[h!]
\begin{center}
\includegraphics[width=0.7\columnwidth]{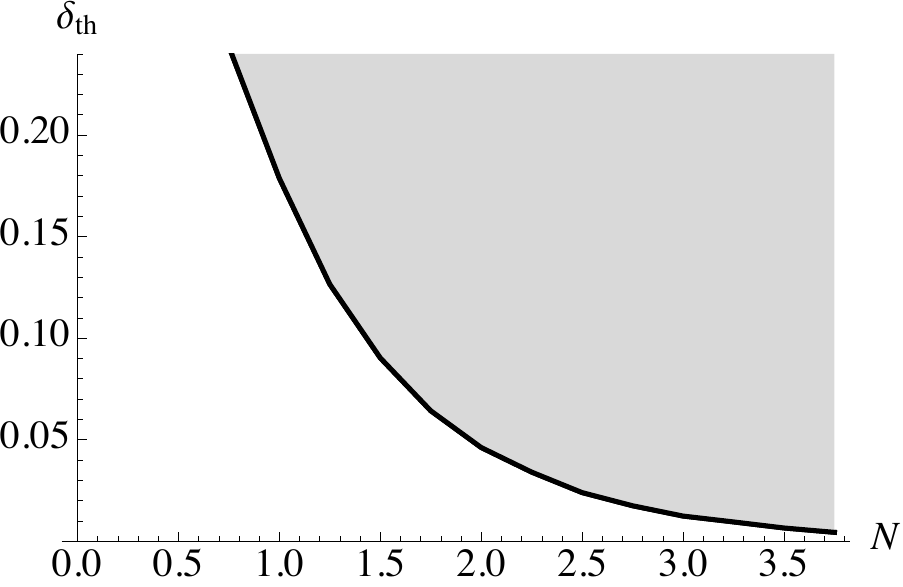}
\end{center}
\caption{(color online) Threshold value $\delta_{\rm th}$ of the 
diffusion coefficient $\delta$ as a function of the signal energy $N$.
The meaning of the threshold value is the following: if $\delta
  \ge \delta_{\rm th}$, then homodyne receiver shows a smaller error
  probability than Kennedy receiver (gray region), the opposite
  otherwise  (white region).
  \label{f:delta:th}}
\end{figure}

This behavior is illustrated in Fig.~\ref{f:delta:th}, where we plot
the threshold $\delta_{\rm th}\equiv\delta_{\rm th}(N)$ on the
diffusion coefficient $\delta$ as a function of the the signal energy
$N$: at any fixed value of $N$ if $\delta \ge \delta_{\rm th}$ then
homodyne receiver exhibits a smaller error probability than Kennedy
receiver in the same experimental conditions, in agreement with Eqs.
(\ref{asyKD}) and (\ref{asyHD}).
\par
{\em Conclusions} --- 
{In conclusion, we have addressed PSK binary optical communication 
in the presence of phase diffusion, i.e., a detrimental noise for
schemes based on coherent signals. We demonstrated experimentally 
that a discrimination strategy based on homodyne detection is robust 
against this kind of noise.} In addition, we have also demonstrated that 
homodyne receivers beat the performances of Kennedy receivers as far 
as the noise is larger than an energy-dependent threshold. Finally,
we have shown that homodyne receivers achieve the Helstrom bound 
on the error probability in the limit of large noise {and for any
  value of the signal energy.}
Our results help to clarify the fundamental limits of quantum 
communications, and show that receivers that perform near the quantum
limit in noisy conditions are realizable with current technology
\cite{Ley12}.
%%%%%%
\section*{Acknowledgments}
This work has been supported by the MIUR project FIRB-LiCHIS-
RBFR10YQ3H.  MGAP thanks F. Ticozzi and P. Villoresi for discussions.
%%%%%%%

%%

\begin{thebibliography}{99}
\bibitem{hel:76} C.~W.~Helstrom, {\it Quantum Detection and Estimation
Theory} (New York: Academic, 1976).
\bibitem{r1}
J. A. Bergou, U. Herzog and M. Hillery, Lect. Not. Phys. {\bf 649}, 415 (2004)
\bibitem{r2}
A. Chefles, Lect. Not. Phys. {\bf 649}, 465 (2004)
\bibitem{r3}
J. A. Bergou, J. Mod. Opt. {\bf 57}, 160 (2010).
\bibitem{Hir96} M. Osaki, M. Ban and O. Hirota, Phys. Rev. A {\bf 54}, 1691
(1996).
\bibitem{Hir99} K. Kato, M. Osaki, M. Sasaki and O. Hirota, 
IEEE Trans. Comm. {\bf 47}, 248 (1999).
\bibitem{lau:06} C.-W.~Lau, V.~A.~Vilnrotter, S.~Dolinar,
  J.~M.~Geremia and H.~Mabuchi, IPN Progr.  Rep. {\bf 42-165}, 1
  (2006).
\bibitem{ken:73} R. S. Kennedy, MIT RLE Quart. Progr.  Rep. {\bf 108}, 1973, p. 219.
\bibitem{tak03} M. Takeoka, M. Ban and M. Sasaki, Phys. Lett. A {\bf 313},
16 (2003).
\bibitem{dol:73} S. Dolinar, MIT RLE Quart. Progr.  Rep. {\bf 111},
1973, p. 115.
\bibitem{ass11} A. Assalini, N. Dalla Pozza and G. Pierobon, 
Phys. Rev. A {\bf 84}, 022342 (2011).
\bibitem{vil12} V. A. Vilnrotter, IPN Progress Report {\bf 42-189}, 1 
(2012).
\bibitem{coo:07} R.~L.~Cook, P.~J.~Martin and J.~M.~Geremia, Nature
  {\bf 466}, 774 (2007).
\bibitem{gre12} M. Gregory, F. Heine, H. Kampfner, R. Lange, M. 
Lutzer and R. Meyer, Opt. Eng. {\bf 51}, 031202 (2012).
\bibitem{ban97} M. Ban, J. Mod. Opt. {\bf 44}, 1175 (1997).
\bibitem{oli:04} S. Olivares and M. G. A. Paris, J. Opt. B: Quantum
  Semiclass. Opt. {\bf 6}, 69 (2004).
\bibitem{car11} G. Cariolaro and G. Pierobon, IEEE Trans. Comm {\bf 58},
623 (2010).
\bibitem{witt:10} C.~Wittmann, U.~L.~Andersen, M.~Takeoka, D.~Sych and
  G.~Leuchs, Phys. Rev. A {\bf 81}, 062338 (2010).
\bibitem{tak08} M. Takeoka and M. Sasaki, Phys. Rev. A {\bf 78} 022320
(2008).
\bibitem{Mig11}
F. E. Becerra, J. Fan, G. Baumgartner, S. V. Polyakov, J.
Goldhar, J. T. Kosloski and A. Migdall, Phys. Rev. A {\bf 84}, 062324
(2011).
\bibitem{Mul12}
C. R. Muller, M. A. Usuga, C. Wittmann, M. Takeoka, Ch. Marquardt, U. L.
Andersen and G. Leuchs, New J. Phys. {\bf 14}, 083009 (2012).
\bibitem{Isu12}
S. Izumi, M. Takeoka, M. Fujiwara, N. Dalla Pozza, A. Assalini, K. Ema and 
M. Sasaki, Phys. Rev. A {\bf 86}, 042328 (2012).
\bibitem{Nai12}
 R. Nair, B. J. Yen, S. Guha, J. H. Shapiro and S. Pirandola, Phys. Rev. A
 {\bf 86}, 022306 (2012).
\bibitem{GardQN} C.~Gardiner and P.~Zoller, {\it Quantum Noise} (Springer, 2004)
\bibitem{geno:11} M. G. Genoni, S. Olivares and M. G. A. Paris, 
Phys. Rev. Lett. {\bf 106}, 153603 (2011).
\bibitem{bri:12} M. G. Genoni, S. Olivares, D. Brivio, S. Cialdi,
 D. Cipriani, A. Santamato, S. Vezzoli and M. G. A. Paris,
 Phys. Rev. A {\bf 85}, 043817 (2012).
\bibitem{Ley12} J. A. Lopez Leya, A. Arvizu Mondragon, E. Garcia, F. J.
Mendieta, E. Alvarez Guzman and P. Gallion,  Opt. Eng. {\bf 51}, 105002
(2012).
\end{thebibliography}
\end{document}